\documentclass[letterpaper, 10 pt, conference]{ieeeconf}  

\IEEEoverridecommandlockouts                              

\usepackage{amsmath,graphicx}
\usepackage{lipsum}
\usepackage{multicol, soul, color, xcolor}

\usepackage{comment}
\usepackage[ruled,vlined]{algorithm2e}
\usepackage{mathtools}
\usepackage[noend]{algpseudocode}
\usepackage[normalem]{ulem}			
\usepackage[utf8]{inputenc}			
\usepackage{cleveref}
\let\labelindent\relax
\usepackage{enumitem}
\usepackage{multirow, hhline}
\usepackage{cite}
\newcommand{\algo}{RoSS}
\newcommand{\minus}{\vspace{-0.05in}}
\newcommand{\new}{\vspace{0.05in}}

\newcommand*{\vertbar}{\rule[-0.5ex]{0.2pt}{2.5ex}}

\title{\LARGE \bf
RoSS: Utilizing Robotic Rotation for Audio Source Separation
}

\author{Hyungjoo Seo, Sahil Bhandary Karnoor, and Romit Roy Choudhury
\thanks{All authors are with the Department of Electrical and Computer Engineering, University of Illinois at Urbana-Champaign, Urbana, IL 61801, USA {$\{hseo17, sahilb5, croy\}$\tt\small @illinois.edu}}%
}

\begin{document}

\maketitle
\thispagestyle{empty}
\pagestyle{empty}

\begin{abstract}
This paper considers the problem of audio source separation where the goal is to isolate a target audio signal (say Alice's speech) from a mixture of multiple interfering signals (e.g., when many people are talking).
This problem has gained renewed interest mainly due to the significant growth in voice controlled devices, including robots in homes, offices, and other public facilities.
Although a rich body of work exists on the core topic of source separation, we find that robotic motion of the microphone --- say the robot's head --- is a complementary opportunity to past approaches. 
Briefly, we show that rotating the microphone array to the correct orientation can produce desired {\em aliasing} between two interferers, causing the two interferers to pose as one.
{\em In other words, a mixture of $K$ signals becomes a mixture of $(K-1)$}, a mathematically concrete gain.
We show that the gain translates well to practice provided two mobility-related challenges can be mitigated.
This paper is focused on mitigating these challenges and demonstrating the end-to-end performance on a fully functional prototype. 
We believe that our {\em Rotational Source Separation} module ({\algo}) could be plugged into actual robot heads, or into other devices (like Amazon Show) that are also capable of rotation.
\end{abstract}

\section{INTRODUCTION}
\label{sec:intro}

As speech recognition and conversational AI matures, voice interactions with robots will become even more popular.
Robots at homes, hospitals, restaurants, airports will all be able to converse with humans.
In this context, recognizing the human's speech/voice is critical, especially since many of these interactions will be happening in noisy environments.
In signal processing, this problem has been called ``source separation'', referring to the ability to separate a voice signal from a mixture of multiple signals.
Source separation has been studied extensively and today's results are impressive, to the extent that $K$ source signals can be separated using $M$ microphones, even when $K$ is slightly larger than $M$.
Observe that the $K>M$ problem is particularly challenging not only because the $K$ sources are unknown, but also because the $K$ channels (over which the signals arrive to the microphones) are also unknown.
Hence, this problem is specifically known as {\em under-determined blind source separation} (UBSS).
\new 

A rich body of work has concentrated on the UBSS problem and today's techniques range from unsupervised methods (e.g., ICA, IVA, Adaptive Beamforming (ABF)), to speech specific techniques (e.g., DUET, Bayesian-DUET), to compressed sensing and supervised deep learning approaches \cite{Wiley_ICA,TASLP07_IVA, Frost_LCMV, DUET04,CompressedSens,ILRMA_sawada_ono_kameoka_kitamura_saruwatari_2019,GatedNN}.
However, for UBSS problems, majority of past works rely on interpolations and regressions since source signal information is lost during the mixing process due to under-determined nature. Therefore, their performance degrades, understandably, as $K$ increases for a given $M$. 
Said differently, any reduction in the $(K-M)$ gap can immediately improve the quality of signal separation.
\new 

\begin{figure}[hbt]
  \centering 
  \includegraphics[width=\columnwidth]{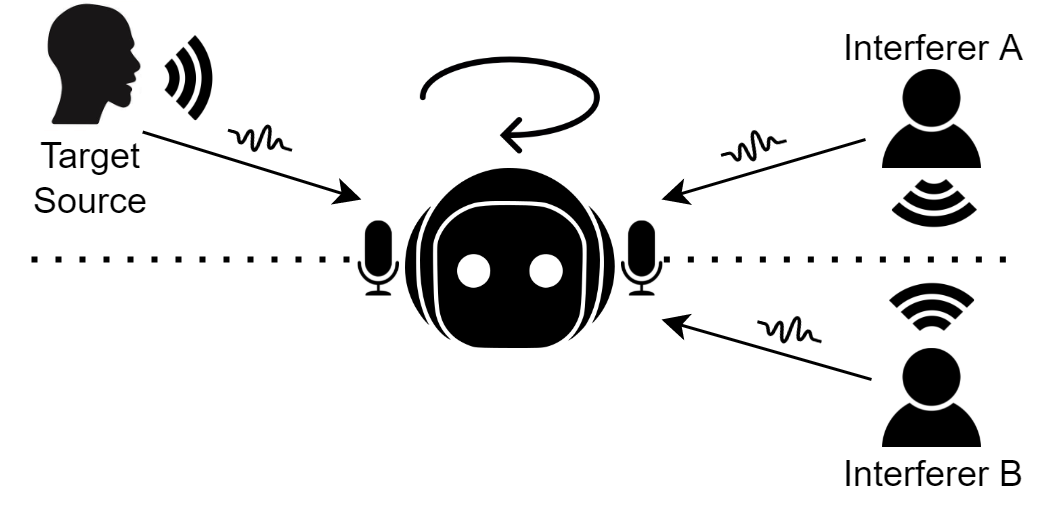}
\minus \minus \minus \minus \minus \minus 
  \caption{Rotation of the microphone array to the correct orientation (i.e., bisecting the source signals) produces the desired spatial aliasing.}
  \label{fig:Ross_idea}
\end{figure}

This paper proposes to leverage robotic mobility to reduce the gap between $K$ and $M$.
At a high level, we intend to rotate a microphone array to an orientation $\theta^*$ such that two interfering signals appear as one interference in this orientation.
The intuition is rooted in {\em angular aliasing}, where two signals arriving from completely different directions will produce the same relative delays at the microphone array, if the array is oriented in the correct angle.
This correct orientation occurs when the line joining the microphones bisects the two interferers as shown in Figure \ref{fig:Ross_idea}.
Thus, deliberate angular aliasing transforms a $[K$=$3$, $M$=$2]$ system into a $[K$=$2$, $M$=$2]$ system, making it solvable.
With many more sources in the real world, reducing $K$ to $K-1$ also offers a clear improvement.

\begin{figure*}
  \centering \includegraphics[width=\textwidth]{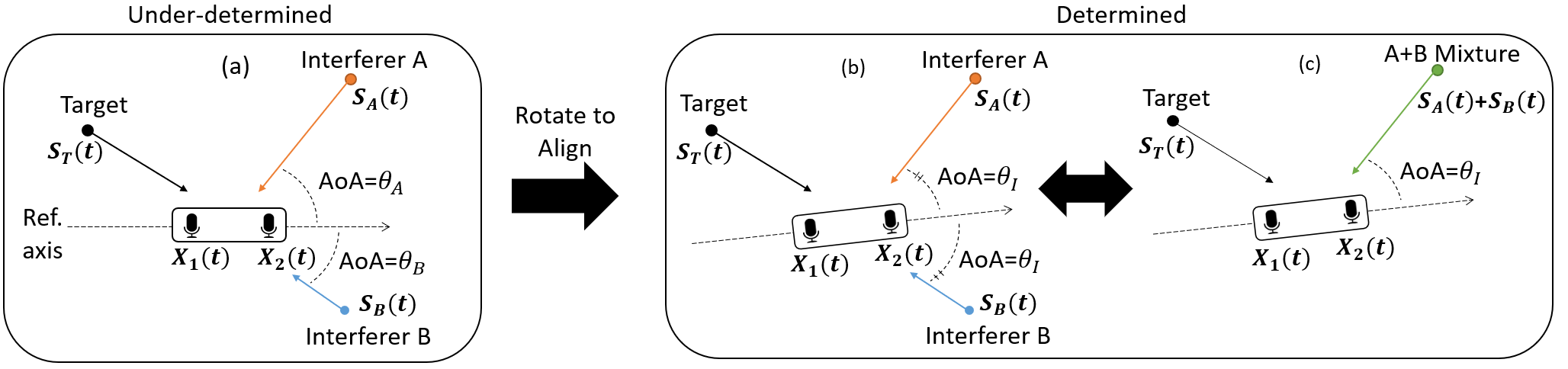}
  \caption{(a) 2-microphone array faced with 3 sources resulting in a UBSS problem. (b) Rotation causes interferers to arrive over the same absolute AoA angle ($\theta$ and $-\theta$). (c) The steering vector for interferers get aliased, resulting in a determined system. }\label{fig:Principle}
\end{figure*}

Separating sources via angular aliasing presents 2 challenges:

\begin{enumerate}
\item Since the angle of arrivals (AoA) of the $K$ signals are not known, the correct microphone orientation $\theta^*$ is unknown as well.
Estimating all $K$ AoAs is difficult with $M<K$ microphones, and worse, AoA estimates are plagued by front-back ambiguities (i.e., it is difficult to tell whether a signal is arriving from a direction $\theta_1$ in front, or $-\theta_1$ from the back).

\item Even if the $K$ AoAs are estimated, it is not clear which interferers should be aligned to maximize source separation performance.
In fact, should the optimal orientation $\theta^*$ always bisect two interferers? 
Or could $\theta^*$ partially align multiple interferers in a way that maximizes separation?

\end{enumerate}

This paper addresses these two problems in Section \ref{sec:Alignment} through a mobility-guided time-frequency method, and then combines with established source separation (SS) techniques to achieve improved performance.
Our {\algo} algorithm is complementary, hence compatible, with all SS techniques.
We implement {\algo} on a rotating microphone prototype (pretending it to be a robot-head), and perform experiments in simulated and uncontrolled environments (Section \ref{sec:Evaluation}).
Results show that {\algo} can achieve over $15$dB of scale invariant signal distortion ratio (SI-SDR) and SI-SDRi \cite{Def_SISDR,TrainigNoisy}, consistently outperforming existing UBSS/BSS methods.
We believe {\algo} could also be effective with smartphones, earbuds, moving video-conference systems, and surveillance cameras, all of which have limited number of microphones but contain actuators or inertial measurement units (IMUs) for angular rotation and sensing.

\section{Formulation and Opportunity}
\label{sec:Formulation}

\subsection{Signal Model}
Let $S_{T}(t),S_{A}(t),S_{B}(t)$ be 3 source signals, of which $S_T$ is the target and others are interference (Fig. \ref{fig:Principle}(a)).
A linear $2$-microphone array receives the mixture of these signals as 
$X_{1}(t)$ and $X_{2}(t)$ and we designate $X_{1}(t)$ as the reference for relative delay calculations.
The signals travel from the far-field over AoAs $\theta_{k}$ (k=T,A,B).
We explain our proposed method with $K=3$ signals and consider $K >3$ later.
\new 

\noindent\textbf{We make the following Assumptions:} \\
(A1) The sound sources are human speech, widely assumed to be mutually independent, non-Gaussian signals.  \\
(A2) Once a speech has been separated, it is possible to tell if it is from the target user (i.e., a voice fingerprint is available). \\
(A3) Sources are not moving in the time scale of seconds.
\new 

Thus, the received (convolutive) signal mixture is: 
\begin{equation}
    X_{1}(t) = \underset{k=A,B,T}\sum  S_{k}(t),\;\; X_{2}(t) = \underset{k=A,B,T}\sum  S_{k}(t+\tau_k)
\end{equation}
where $\tau_k = \frac{d}{v_p}cos(\theta_k)$, (k=T,A,B), are time-difference-of-arrivals (TDOA) between the microphones, while $v_p$ and $d$ denote sound propagation speed and spacing between microphones, respectively.
\new 

Thus, in the time-frequency domain:
\begin{equation}\label{eq:STFT}
    \Vec{X}(f) = \Vec{a}_{T}(f)S_{T}(f) + \Vec{a}_{A}(f)S_{A}(f) + \Vec{a}_{B}(f)S_{B}(f) 
\end{equation}

In matrix form, this equation becomes:
\begin{align}
\begin{bmatrix}
X_1(f) \\ 
X_2(f) \\
\end{bmatrix}
=
\begin{bmatrix}
\vertbar    & \vertbar      & \vertbar \\
\Vec{a}_T   & \Vec{a}_A    & \Vec{a}_B \\
\vertbar    & \vertbar      & \vertbar \\
\end{bmatrix}
\begin{bmatrix}
S_T(f) \\ S_A(f) \\ S_B(f) 
\end{bmatrix}
\label{eq:A_decomp}
\end{align}

where $\Vec{a}_k$ = $[1 \; \text{exp}(j2\pi f \tau_k)]^{T}$ (k=A,B,T) is the steering vector.
Note that even if all $\Vec{a}_k$'s are known, the system is still under-determined.

\subsection{Interference Alignment}\label{subsec:alignment}

What if we rotate the array such that the line joining the microphones bisect the two interferers? 
While the correct rotation angle needs to be inferred blindly, for now let us assume we know it.
Fig.\ref{fig:Principle}(b) shows the outcome. 
Since the new AoAs of the two interferers are now $\theta_I$ and $-\theta_I$, their corresponding TDOAs become equal, or aliased, as follows:
$$\tau'_A = \frac{d}{v_p}cos(\theta_I)=\frac{d}{v_p}cos(-\theta_I)=\tau'_B$$
Thus, in frequency domain, interferers A and B have identical array vectors $\Vec{a}_I(f) = [1 \; \text{exp}(j2\pi f \tau_I)]^{T}$ where $\tau_I=\tau'_A=\tau'_B$.
Hence, the new measurement vector $\Vec{X}'(f)$ is:
\begin{align}
\begin{bmatrix}
{X_1}^{'}(f) \\ 
{X_2}^{'}(f) \\
\end{bmatrix}
=
\begin{bmatrix}
\vertbar    & \vertbar       \\
\Vec{a}_T   & \Vec{a}_I     \\
\vertbar    & \vertbar       \\
\end{bmatrix}
\begin{bmatrix}
S_T(f) \\ S_A(f) + S_B(f) 
\end{bmatrix}
\label{eq:A_merged}
\end{align}
This expression means that the array {\em would sense two groups of signals, not three} (i.e., the target and the sum of two interferers).
Fig. \ref{fig:Principle}(c) shows the two signals arriving from distinct angles.
This produces a determined system of equations except that one of the mixed signals arriving from AoA $\Vec{a}_I$ is actually a sum of independent sources.
If the sum $(S_{A}(f) + S_{B}(f))$ is independent of the target signal $S_{T}(f)$ (as shown next), we can apply classical source separation.

\subsection{Sum of Mutually Independence Sources}
We briefly show that a mixture of two independent sources remains independent from the third source when all three are mutually independent. 
Define $A$, $B$ and $T$ as mutually independent continuous random variables, and $J = A + B$ is a fourth random variable.
Let $F_i(\cdot)$ and $f_i(\cdot)$ be CDF and PDF of variable i, respectively.
Then, the joint distribution of $T$ and $J$ can be written as:
\begin{equation} \label{eq:proof}
     \begin{split}
         F_{JT}(j,t)&=P(A+B\leq j, T\leq t) \\
         &= \int P(A+B\leq j, T\leq t|A=a)f_A(a)\,da \\
         &= \int P(B\leq j-a, T\leq t)f_A(a)\,da \\
         &= \int P(B\leq j-a)f_A(a)\,da \cdot P(T\leq t) \\
         &= P(A+B\leq j) \cdot P(T\leq t) 
         = F_J(j)F_T(t) 
     \end{split}
 \end{equation}
Therefore, $J$ and $T$ are also mutually independent \cite{book_stat}.

\section{Alignment by rotational motion}
\label{sec:Alignment}
Our goal now is to correctly rotate the microphone so that interferers get aligned. 
For this, we first need to estimate the source AoAs, and then determine the correct microphone rotation as a function of these estimated AoAs.

\subsection{Estimating AoAs in Under-determined Scenarios}
\begin{figure}[hbt]
\centering 
\includegraphics[width=\columnwidth]{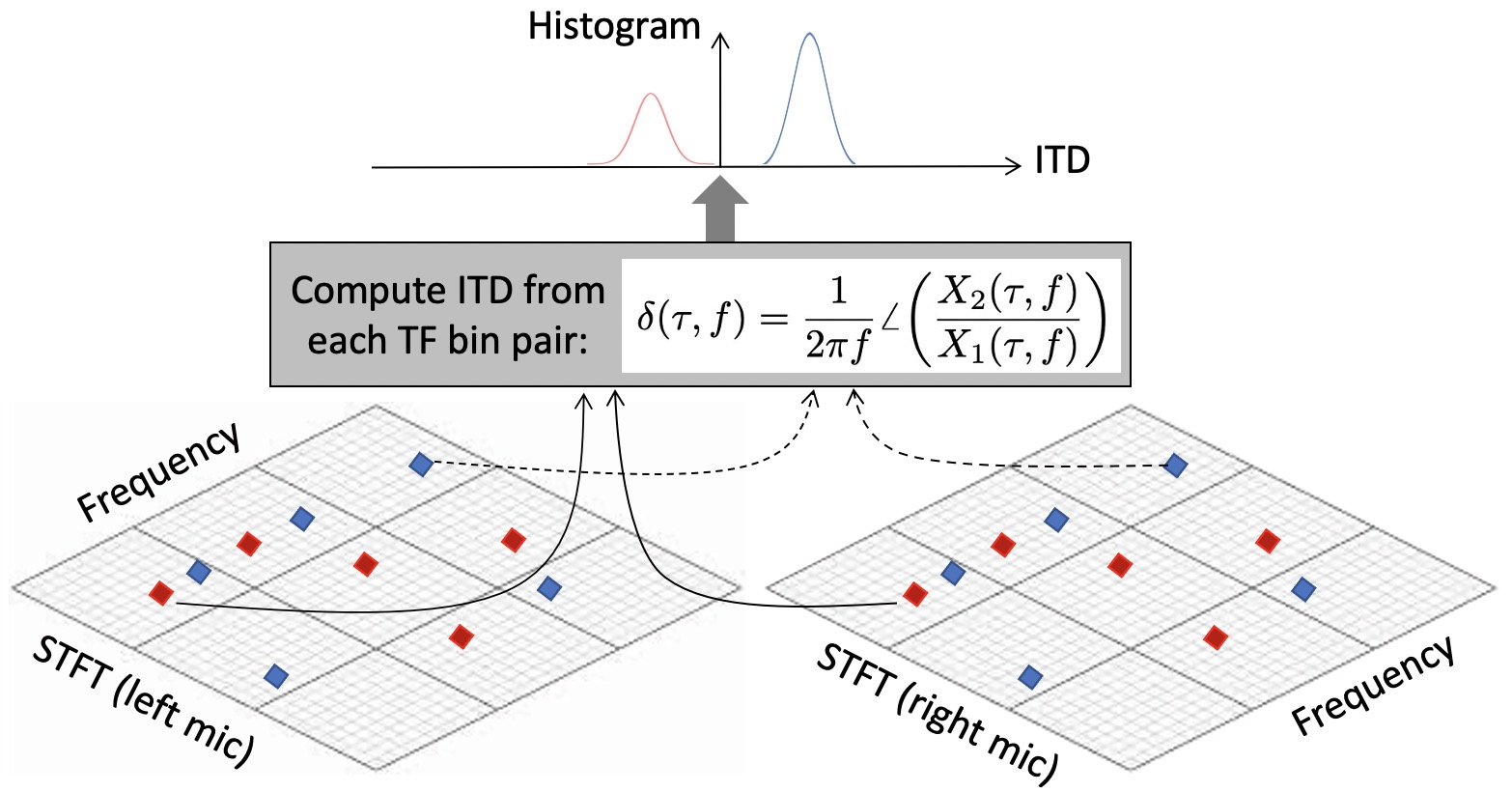}
\caption{ITD computed from TF bins produce 2 clusters around two mean ITDs. These mean ITDs are estimates of AoA.}
\label{fig:duet}
\end{figure}

Estimating $K$ AoAs with $M<K$ microphones is known to be a hard problem for general signals.
However, literature has shown promise with speech signals due to what is known as the {\em W-Disjoint Orthogonality} (WDO) property \cite{DUET04}.
Briefly, extensive experiments have shown that speech from two humans have a low probability of collision in a given time-frequency (TF) bin.
Thus, if one calculates the TDOA for each TF bin --- called {\em inter-microphone time difference} (ITD) --- one can extract information about AoAs.
Fig. \ref{fig:duet} illustrates this with a toy example of red and blue signals; the calculated ITDs from the red and blue TF bins form $2$ clusters.
The means of these clusters partly reveals the red/blue signal's AoA.

\begin{figure}[hbt]
\centering 
\includegraphics[width=\columnwidth]{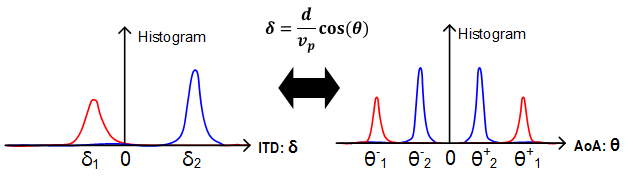}
\minus \minus \minus \minus \minus \minus
\caption{2 ITD clusters gets mapped to 4 clusters in $(-\pi, \pi]$ AoA space.}
\minus \minus
\label{fig:mapping}
\end{figure}

Unfortunately, the mapping between ITD and AoA is not $1$:$1$ because AoAs of both $\theta$ and $-\theta$ produce identical ITDs at the microphone array. In Fig. \ref{fig:mapping}, see how $2$ clusters get mapped to $4$ AoAs (of which 2 AoAs are spurious). This is classically known as the {\em front back ambiguity}.
Rotating the microphone to the correct orientation $\theta^*$ would obviously need to resolve this ambiguity first.


\subsection{Rotation Resolves AoA Ambiguity}
\begin{figure}[hbt]
  \centering \includegraphics[width=\columnwidth]{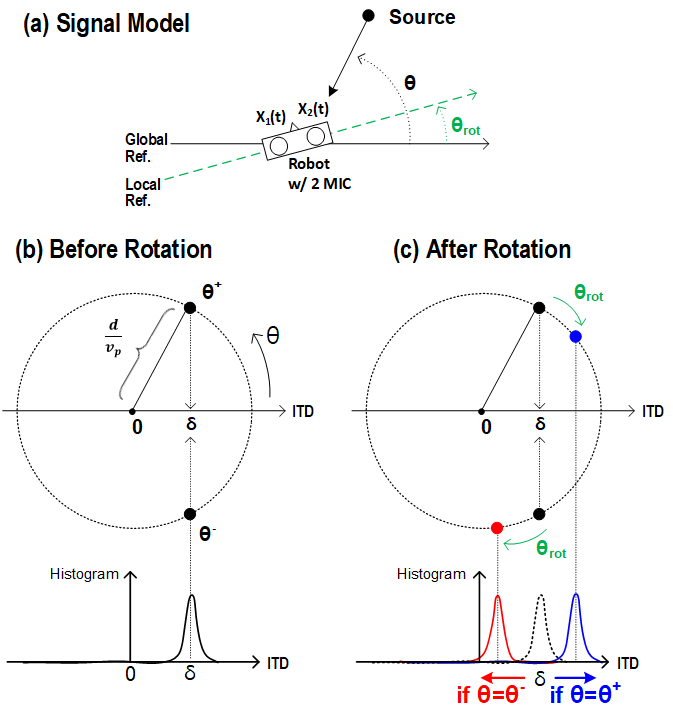}
\minus \minus \minus \minus \minus
  \caption{(a) 2-microphone array rotating by $\theta_{rot}$ when faced with a source. (b) Front-back ambiguity in AoA. (c) Rotational motion is equivalent to all sources rotate by $\theta_{rot}$ opposite direction, and AoA ambiguity disappears.}\label{fig:Rot_Split}
\end{figure}

To resolve AoA ambiguity, we propose to rotate the microphone and observe the change in the ITDs.
Depending on the ITD change (positive or negative) it is possible to resolve whether the source's AoA is in front ($\theta$) or back ($-\theta$).
Fig. \ref{fig:Rot_Split}(a) illustrates an example counter-clockwise rotation of 
$\theta_{rot}\in [0,180)$.
From the microphone's reference frame, the source AoAs rotate in the clockwise direction.
If the original AoA was in front ($\theta$), then Fig. \ref{fig:Rot_Split}(c) shows how the new ITD increases (i.e., a right shift in the ITD axis), and vice versa when the AoA is $-\theta$.

Equation \ref{eq:ITD-AoA_new} and \ref{eq:ddelta} show this analytically, where $\delta'$ is the ITD after rotation, and $\Delta\delta'$ is change in ITD after the rotation. 
 \begin{equation}\label{eq:ITD-AoA_new}
    \delta' =\frac{d}{v_p}cos(\theta-\theta_{rot}) 
\end{equation} 
 \begin{equation}
 \begin{split}\label{eq:ddelta}
    \Delta\delta' &= \delta' - \delta \\
    &=\frac{d}{v_p}\big(cos(\theta-\theta_{rot})-cos(\theta)\big) \\
    &=\frac{2d}{v_p}sin(\frac{\theta_{rot}}{2})sin(\theta-\frac{\theta_{rot}}{2})
 \end{split}
\end{equation} 

As discussed earlier, the sign of $\Delta\delta'$ is different based on whether the source is in front or back.
This sign offers a reliable feature to resolve the front-back ambiguity. 
Of course, challenges emerge in real scenarios, discussed next.
\new

\begin{figure}[hbt]
  \centering \includegraphics[width=\columnwidth]{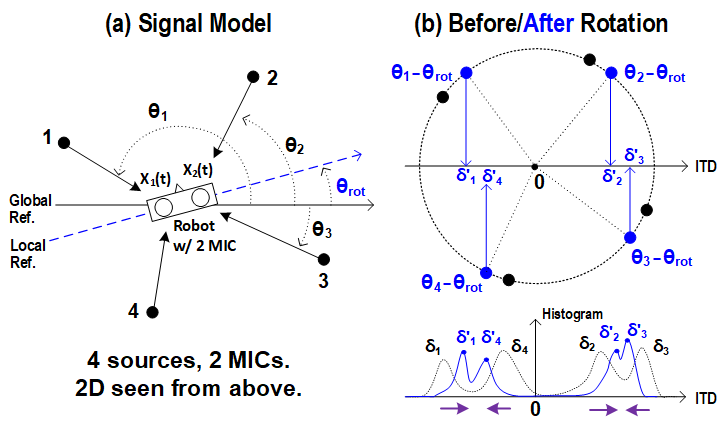}
  \caption{(a) 2-microphone array rotating by $\theta_{rot}$ when faced with K=4 sources. (b) Rotation shifts ITD values of K source accordingly.}\label{fig:Rot_K}
\end{figure}

What happens when real situations have more than one source? 
Consider $K=4$ sources in Fig. \ref{fig:Rot_K}(a). 
For easier notations, let us map the each signal as $k$=$\{1,2,3,4\}$.
Thus, if the ITD was $\delta_k$ before rotation, then after rotation, the ITD becomes:
 \begin{equation}\label{eq:ITD-AoA_new_k}
    \delta'_k =\frac{d}{v_p}cos(\theta_k-\theta_{rot}) 
\end{equation} 
where $\theta_k$ was the true global AoA value before rotation.

 As shown in Fig. \ref{fig:Rot_K}(b), the sign based feature becomes challenging as the peaks from $\delta'_k$ and $\delta_k$ pairs begin to merge in a crowded histogram.
 The situation is worse when K becomes larger or when the measurement is noisy. 
 This motivates our following approach to identify K AoAs using $\theta_{rot}$ and 2 ITD measurements.

\new

\noindent{\bf Hypothesis Testing:}

From a current estimate of ITD, and after a rotation of $\theta_{rot}$, we can derive two expected ITDs as:
 \begin{equation}\label{eq:ITD-est}
    \hat{\delta'}_k =\frac{d}{v_p}cos(\hat{\theta}_k-\theta_{rot}) 
\end{equation} 

where $\hat{\theta}_k$ = $\{ +\hat{\theta}_k$, $-\hat{\theta}_k \}$ still has front-back ambiguity.
From now on, let us call $\{ +\hat{\theta}_k$, $-\hat{\theta}_k \}$ as $\hat{\theta}_{k,p}$, $\hat{\theta}_{k,n}$.

\begin{figure}[hbt]
  \centering \includegraphics[width=\columnwidth]{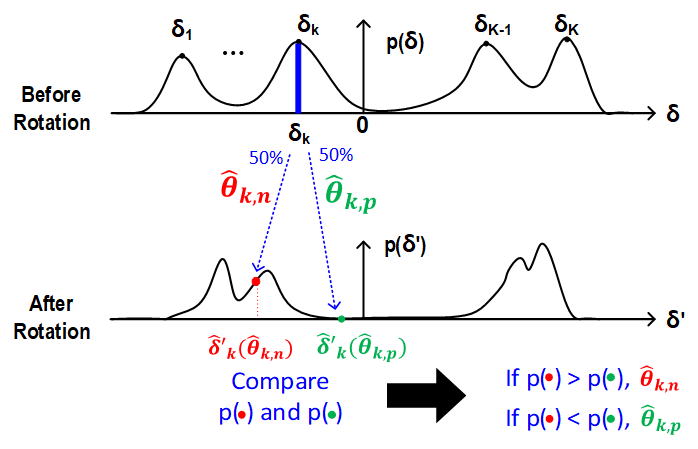}
  \caption{Binary hypothesis testing on estimation of K AoA values using rotation angle.}\label{fig:Binary_HT}
   
\end{figure}

We now apply binary hypothesis testing by comparing the expected $\hat{\delta'}_k$ with the measured ITD histogram $\delta'_k$ obtained after rotation.
This also reveals the more likely value of $\theta_k$, essentially a maximum likelihood estimator (MLE).
Said differently, the probability density function (PDF) of $\delta'_k$ near one of the $\hat{\delta'}_k$ values is expected to be higher than the counterpart as shown in Fig. \ref{fig:Binary_HT}.
This gives us the correct AoA as:
 \begin{equation}\label{eq:Hypothesis}
    \theta_k^{\ast} = \hat{\theta}_{k,i} 
\end{equation} 
where
 \begin{equation}\label{eq:Hypothesis}
 \begin{split}
 i &= \underset{i=p, n}{\text{argmax}} \: p(\delta'|\hat{\delta'}_k(\hat{\theta}_{k,i}))\\
 &=\underset{i=p, n}{\text{argmax}} \: \frac{p(\delta'=\hat{\delta'}_k(\hat{\theta}_{k,i}))}{p(\hat{\delta'}_k(\hat{\theta}_{k,i}))}\\
 &=\underset{i=p, n}{\text{argmax}} \: p(\delta'=\hat{\delta'}_k(\hat{\theta}_{k,i}))
 \end{split}
\end{equation} 

where p($\delta'$) is a PDF of $\delta'$ that can be acquired by fitting the measured new ITD histogram $\delta'$, and p($\hat{\delta'}_k$) is a prior distribution on $\hat{\theta}_{k,i}$.
We use equal priors, 1/2, for each $i$ in Equation \ref{eq:Hypothesis}. 
Kernel density estimation (KDE) can be used to fit the distribution based on the histogram acquired by ITDs.
By repeating the estimation K times for $k=1,...,K$, we estimate $K$ $\theta_k^{\ast}$ for $K$ sources.

 In practice, however, several more issues affect the above-mentioned solution.
 
\noindent$\bullet$ There is no guarantee that all $K$ ITD values would be prominent in first measurement because some source pairs can be image pairs, i.e., AoAs of $30^\circ$ deg and -$30^\circ$, respectively. 

\noindent$\bullet$ p($\delta'$) may not be higher in the expected region because other signals can appear in the opposite region after rotation, resulting in an erroneous decision in Equation \ref{eq:Hypothesis}.


\noindent$\bullet$ Since the $\delta'$ to $\theta$ mapping is non-linear, the variance of a peak in ITD axis may not be preserved when mapped to its AoA. 
Thus, AoA estimation error varies depending on true AoA values at each measurement. 
For instance, AoA values near 0 or 180 result in more estimation errors than those in 90 or -90 degrees for the same variances in the ITD domain.

To overcome these problems, let us consider taking more rotational measurements by rotating R times with $\theta_{rot}^{(r)}$ to get R ITD distributions $p(\delta^{(r)})$ where $r=1,...,R$ and $r=0$ is for no rotation case.
\new 

\noindent{\bf Markov Model Testing:} \\
After the r-th rotation, if $K^{(r)}$significant peaks are found in $p(\delta^{(r)})$ at $\delta^{(k)}_k$, where $k=1,...,K^{(r)}$, then local AoA values at the r-th rotation for $K^{(r)}$ sources is estimated as:

 \begin{equation}\label{eq:theta_rth}
    \theta^{(r)\ast}_k = \hat{\theta}^{(r)}_{k,i} 
\end{equation} 
where
 \begin{equation}\label{eq:MLE_rth}
 \begin{split}
 i &= \underset{i=p, n}{\text{argmax}} \: p(\delta^{(r)}|\hat{\delta}_k^{(r)}(\hat{\theta}^{(r-1)}_{k,i}))\\
 &=\underset{i=p, n}{\text{argmax}} \: p(\delta^{(r)}=\hat{\delta}_k^{(r)}(\hat{\theta}^{(r-1)}_{k,i}))
 \end{split}
\end{equation} 

Since the total rotation after $r$ times is known and equals $\sum_{r'=1}^{r}\theta_{rot}^{(r')} $, the global AoA values from the r-th measurements, $\theta_{k,r}^{\ast}$, can be estimated as:
 \begin{equation}\label{eq:total_rot}
    \theta_{k,r}^{\ast} = \theta^{(r)\ast}_k + \sum_{r'=1}^{r}\theta_{rot}^{(r')}
\end{equation}

Noting that rotational movement only changes a state to the next state, the given process can be viewed as a Markov process where the $(r)$-th state is only affected by the $(r-1)$-th state. Therefore, estimation of Equation \ref{eq:MLE_rth} can be repeated iteratively from $r=1$ to $R$ to generate a total number of $\sum_{r=1}^{R} K^{(r)}$ of $\theta_{k,r}^{\ast}$ values. 
By capturing major clusters within the data of $\{\theta_{k,r}^{\ast}\}$ where $k=1,...,K^{(r)}$ and $r=1,...,R$, mean values of clusters, K global AoA estimates $\theta_k^{\ast}$, can be estimated. 
\new 

Compared to 1-rotation estimator, R-rotation estimator statistically improves the issues mentioned above by relying on multiple measurements taken at different orientations. 
Of course, robustness improves for greater $R$ if time duration at each measurement remains the same. 



\subsection{Optimal Angle for Source Separation}

Once K AoA values $\theta_k^{\ast}$ are estimated, the optimal rotation angle for source separation can be found.

\new

\noindent{\bf $\bf{K\leq3}$ Case:} For $K\leq2$, no rotation is needed for source separation since the system is already (over-)determined and all K sources can be recovered at the same time via techniques such as ICA. For $K=3$, a pair of sources can be aligned as explained in Section \ref{subsec:alignment} to recover remaining source. To enhance source i, rotation angle for alignment $\theta_{i,align}$ is a bisecting angle between two interfering sources.
 \begin{equation}\label{eq:alignment_ang_K3}
    \theta_{i,align} = \frac{1}{2}\sum_{k \neq i}^{K} \theta_k^{\ast} = \frac{(\sum_{k=1}^{K} \theta_k^{\ast})-\theta_i^{\ast}}{2}
\end{equation}

By setting $\theta_{rot}=\theta_{i,align}$, determined source separation technique such as ICA can be applied to recover $S_i(t)$. 

\new

\noindent{\bf $\bf{K>3}$ Case:} Even if a pair of interferers are aligned, the system is still under-determined because an alignment of a pair only reduces effective K by 1, i.e., $K-1$ can still be greater than $M$ where $M=2$. 
Therefore, UBSS techniques that utilize TF masks (e.g., DUET \cite{DUET04}) can be used to separate $K-1$ groups of sources. T-F masking methods estimate masks for each T-F bin based on ITD measurements.
Therefore, source separation performance depends on both clear identification of $\delta_k(\tau,\delta)$ and small number of overlapping T-F bins. 
Importantly, our proposed interference alignment approach provides benefits in both cases.

First, alignment of two nearest angular neighbors of target source yields maximum isolation in $\delta$ domain from rest of the sources as shown in Fig. \ref{fig:Triangles}(a) with few exceptions (b). One way to check whether it's (a) or (b) is to draw all triangles containing the target and nearest angular neighbor sources, and count the number of acute triangles. If there is at least 1 acute triangle as in (a), then aligning nearest neighbors is best isolation angle. Otherwise, it is AoA dependent as in (b) \footnote{Minimum angle difference with adjacent source $\Delta\theta_{min} = \theta_i - \theta_{adj} < 0.68\approx38.94$ deg \& all other interferers $\theta_k \in (-\theta_{adj},\theta_{adj})$ \& $|\theta_k| > cos^{-1}(3cos(\theta_i-\theta_{rot}))+\theta_{rot}$ for $\Delta\theta_{min}>0$. This case is not considered in this paper since the condition is strict.}. Thus, there is optimal rotation angle where $\delta_i$ can be clearly identified in histogram of p($\delta$) without being interfered by adjacent interferers. 

Second, proportion of under-determined T-F bins reduces after interference alignment which allows sparsity-based UBSS techniques \cite{TCASI19_UBSS_Sparse_TF} to recover such T-F bins. Therefore, rotational alignment angle for $K>3$ can be generalized as:

\begin{figure}[hbt]
  \centering \includegraphics[width=\columnwidth]{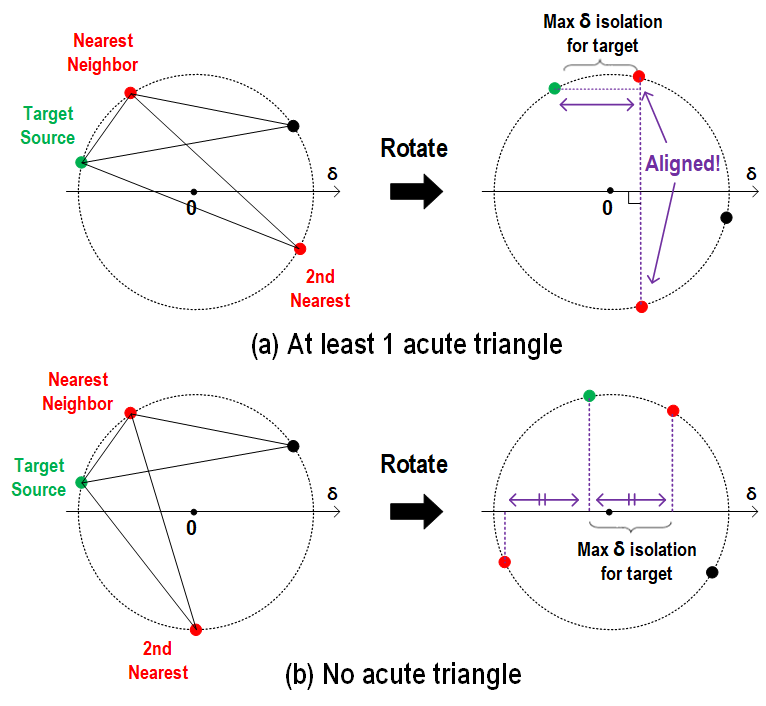}
       
  \caption{In ITD-AoA plot, there are two types of maximum target isolation in $\delta$ domain.(a) Nearest neighbors are aligned. (b) No sources are aligned}\label{fig:Triangles}
   
\end{figure}

 \begin{equation}\label{eq:alignment_ang_K4}
    \theta_{i,align} = \underset{\theta_{rot}}{\text{argmax}} \: \underset{k}{\text{min}} \:|\delta_i-\delta_k|
\end{equation}

\noindent{\bf Target Signal Identification:} 
At each alignment angle, target signal feature can be used to perform voice recognition to assign scores to know whether the target signal is on recovered group or mixture group where K-1 iterations are needed in worst case scenario.

\subsection{{\algo} Algorithm}
Rotational Source Separation (RoSS) algorithm contains two modules: i) Explore and ii) Exploit as shown below:

\noindent$\bullet$ At every rotational angle, STFT is performed on microphone signals after applying LPF to prevent spatial aliasing.

\noindent$\bullet$ RoSS runs explore module first to identify AoAs. Then, it runs exploit module to rotate to right alignment angle for source enhancement. (back-to-back)

\noindent$\bullet$ If any interference signal moves, Score value of voice recognition would drop due to misalignment of interferers. Then, RoSS gets re-initiated from the beginning. Note that movement of target signal does not affect interference alignment much.

\begin{figure}[h]
 
\centering \includegraphics[width=\columnwidth]{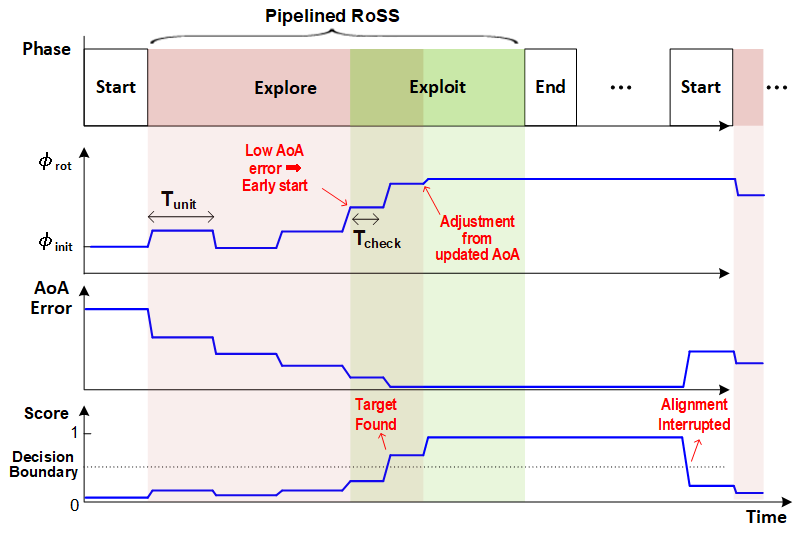}
 \minus \minus \minus \minus 
\caption{\small{System-level demonstration example of pipelined RoSS}}\label{fig:Pipeline}
   
\end{figure}

\noindent$\bullet$ RoSS can be run faster through a pipeline of two modules: Once brief AoA estimate is acquired before exploration ends, move to one of the estimated alignment angle prematurely then use this movement to update the AoA estimation. Repeat until target source is found. This way, both improving AoA accuracy and target source search can be done at once. See Fig. \ref{fig:Pipeline} for detailed operation.

\minus\minus

\begin{algorithm} \caption{Explore Module} \label{algo:Explore}\small
    \SetAlgoLined
    \KwIn{Starting position, $\{\theta^{(r)}_{rot}\}$}
    \KwOut{$\{\theta_k^{\ast}\}, K, \phi_{rot}$}
    Initialize $\phi_{rot} = 0$, $r=0$  \;

     \While{$\phi_{rot} \leq$ 180 [deg]}{
    Fit ITD histogram with KDE to get $p(\delta^{(r)})$ \;
    Find prominent peaks $\{\delta^{(r)}_k\}$ from $p(\delta^{(r)})$ \;
    \If{$r=0$}{
    $r = r + 1$, $\phi_{rot}=\phi_{rot}+\theta^{(r)}_{rot}$, Rotate by $\theta^{(r)}_{rot}$ \;
    continue\;
    }
    From $\{\delta^{(r-1)}_k\}$, $\theta^{(r)}_{rot} \rightarrow$ get $\{\hat{\delta}^{(r)}_k\}$ (\ref{eq:ITD-est}) \;
    From $\{\hat{\delta}^{(r)}_k\}$, $\{\delta^{(r)}_k\} \rightarrow$ get $\{\theta^{(r)\ast}_k\}$ (\ref{eq:theta_rth}) (\ref{eq:MLE_rth})\;
    From $\{\theta^{(r)\ast}_k\} \rightarrow$ $\{\theta_{k,r}^{\ast}\}$  (\ref{eq:total_rot}) \;
    K-Means($\{\theta_{k,r}^{\ast}\}$) w/ inertia $\rightarrow$ $K_r$ cluster means \;
    (break; if $K_r$ does not change for 2 times)

    $r = r + 1$, $\phi_{rot}=\phi_{rot}+\theta^{(r)}_{rot}$, Rotate by $\theta^{(r)}_{rot}$ \;
 }
    Save cluster means to $\{\theta_k^{\ast}\}$, save $K_r$ to K, save $\phi_{rot}$ \;
\end{algorithm}
\minus\minus

\begin{algorithm} \caption{Exploit Module}\label{algo:Exploit}\small
    \SetAlgoLined
    \KwIn{$\{\theta_k^{\ast}\}, K, \phi_{rot}$}
    \KwOut{$\hat{s}_i(t)$ (i of interest)}

    From K, determine $\{\theta_{i,align}\}$ using $\{\theta_k^{\ast}\}$ \;
    J $\leftarrow \underset{i}{\text{Sort}}_{min}^{max}\: |\{\theta_{i,align}\}- \phi_{rot}|$ \;

 \For{j in J}{
Rotate to $\theta_{j,align}$ \;
For small time duration $T_{check}$ [s], \\
\uIf{$K>3$}
{
    Perform DUET to get K-1 signal groups\;    
}
\ElseIf{$K\leq3$}
{
    Perform IVA to get K-1 signal groups \; 
}
VoiceRecognition(K-1 groups) $\rightarrow$ $\{$Scores$\}$\;
\If {max$_i$$\{$Scores$\}>$ Decision Criteria}{break \;}
 }
Perform IVA or DUET for $T_{run}$ [s] to get $\hat{s}_i(t)$ \;
\end{algorithm}

\minus\minus

\section{Evaluation}
\label{sec:Evaluation}
\subsection{Settings}

\noindent{\bf Data Model:}
We test {\algo} on 1-minute-long speech recordings from the LibriTTS dataset\cite{LibriTTS}. We made each signal power to be identical, i.e., $SIR \approx -10log(K-1)$ for K-sources. Also, K sources are selected in a pre-defined AoA order of [170, 70, -50, -120, 10]. For instance, [170, 70, -50] is used for $K=3$ case.

                \begin{table}[ht]  
                \caption{Details of 5 evaluation settings} 
                 
                \centering 
                \begin{tabular}{c c c c} 
                \hline\hline 
                Setting &  1, 2, 3 & 4 & 5 \\
                 \hline\hline
                 Type & Simulation &   Measurement & Measurement  \\
                 Location & Computer &   Outdoor & Indoor  \\
                 SNR [dB] & 15  & 15.4 &23\\
                 $T_{60}$: Reverb Time [ms] & 0, 450, 700  & $\approx$ 0  & N/A\\
                  Room Size [$m\times m$] & $10\times 10$ & $>20\times 20$ & $\approx 6\times8$ \\
                 Distance to speakers [m] & 2.5  & 2 & 2 to 2.5 \\
                $\theta^{(r)}_{rot}$ [deg] & 15 ($\forall  r$)   & 15 ($\forall  r$) & 15 ($\forall  r$) \\
                \end{tabular}
                \label{table:Settings}
                \end{table}

\begin{figure*}[hbt]
 
\centering \includegraphics[width=\textwidth]{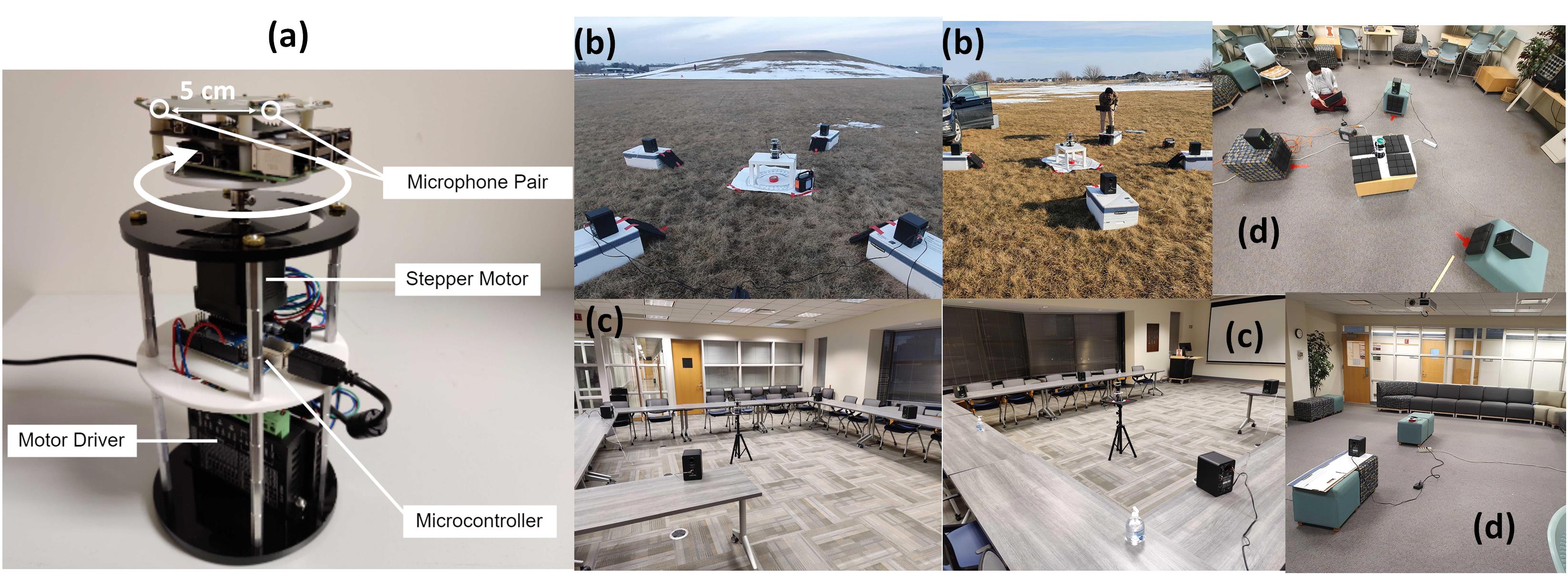}
  \minus \minus \minus \minus \minus
\caption{\small{\small(a) Custom-built rotary platform with MICs. (b) Outdoor field (Setting 4). (c) Conference room (Setting 5). (d) Testing scenes.}}\label{fig:Exp_Pics}
  
\end{figure*}
\minus

\noindent{\bf Experiment Settings:}
We have implemented {\algo} on a custom built rotary platform actuated by a NEMA-17 stepper motor (Fig. \ref{fig:Exp_Pics}(a)). The motor was controlled in open-loop using a TB-6600 motor driver where its peak rotation speed and acceleration is $225$ $deg/s$ and $112.5$ $deg/s^2$ each. 
ReSpeaker microphone array for Raspberry Pi is mounted on the rotary platform and 2 adjacent microphones, with 5cm spacing, are used to record audio signal. Experiments were tested and conducted in more than two environments including one outdoor and two indoor locations as depicted and explained in Fig. \ref{fig:Exp_Pics}(b,c,d) and Table-\ref{table:Settings}. In each environment, the platform was placed at the center and K sources were played from individual speakers placed radially at a distance of $2$ to $2.5$m from the platform.

\new

Algorithms are run in Python with sampling frequency of $16$kHz, STFT frame lengths of 512 or 1024 with 25\% overlap with adjacent frames. For source separation module, we use natural gradient-based IVA \cite{TASLP07_IVA}, DUET \cite{DUET04} and MVDR \cite{Frost_LCMV} for demonstration purpose of our concept. Other state-of-the-art UBSS/BSS techniques \cite{GatedNN,ConvTasNet} could also be adopted for better separation performance. For rotation step $\theta^{(r)}_{rot}$, we used constant rotation by 15 degrees. Rotation can be planned judiciously such as rotating back and forth for instance.

\new

However, in performance evaluation, since it is difficult to compare enhanced target source with the true target source alone in measurement for comparison, we measured $X_1(t)$ twice, with and without interference at different time points. Note that, this would degrade source separation performance naturally due to external noise sources because two acoustic environments can be different at each time point.

\new
\noindent{\bf Simulation Settings:}
In settings 1-to-3 in Table-\ref{table:Settings}, two convolutive mixtures, $X_1(t), X_2(t)$, are generated based on room impulse response (RIR) generator \cite{rir_generator} with following conditions. 

\noindent$\bullet$ Room size: 10m x 10m (2-dimensional space assumed).

\noindent$\bullet$ Two omni-directional identical microphones with spacing of 5 cm are located at the center, rotating around center.

\noindent$\bullet$  Gaussian noise is added so that microphone SNR is 15 dB while maintaining SIR of -10log(K-1) dB

\noindent$\bullet$ Separated sources are evaluated by comparing with each source alone measured at the reference microphone $X_1(t)$.

Algorithm settings are similar as in measurement settings except for $24$kHz sampling frequency.

\subsection{Performance Evaluation}

\begin{figure*}
  \centering \includegraphics[width=\textwidth]{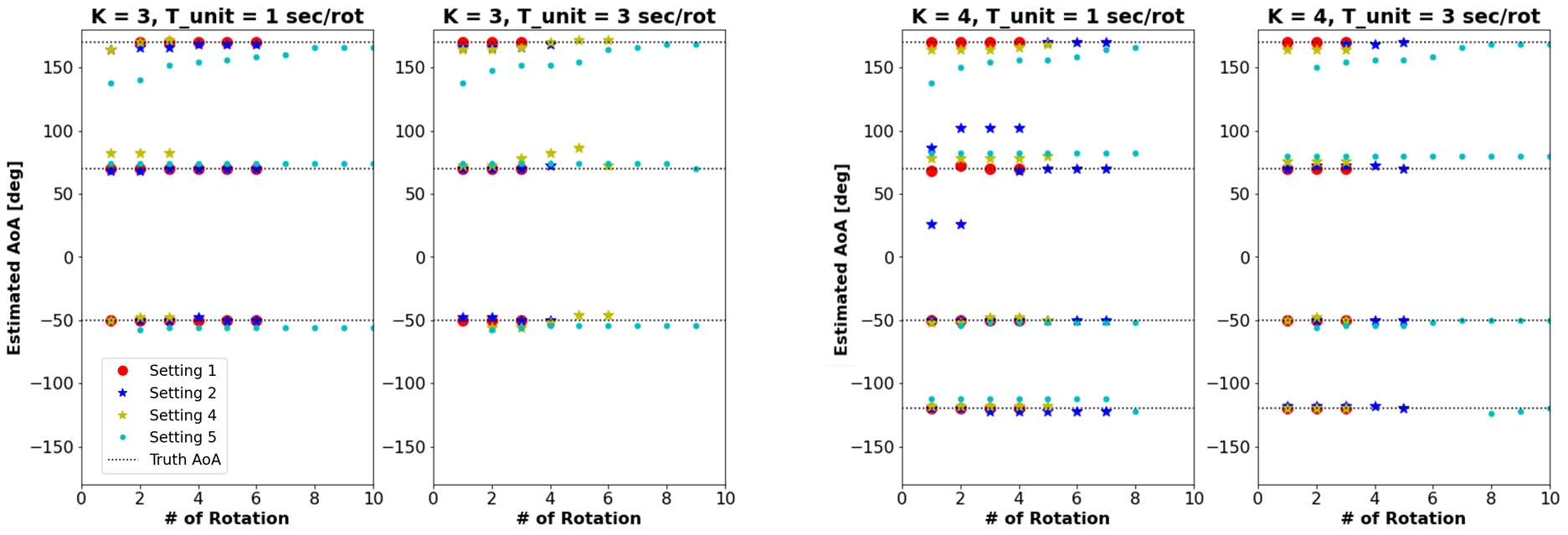}
   \minus \minus \minus \minus \minus \minus     
  \caption{AoA estimation result over time in different settings and different number of sources and unit times. Dotted line represents truth AoAs.}\label{fig:AoA_perf}
   
\end{figure*}

\noindent{\bf AoA estimation:}
Fig. \ref{fig:AoA_perf} shows AoA estimation results based on number of rotational steps in 4 different combinations of K and $T_{unit}$ where K is number of sources and $T_{unit}$ comprises of listening time and moving time per rotation while majority is listening time since it takes less than 0.1 seconds to rotate 15 deg. As claimed in Section \ref{sec:Alignment} and (\ref{eq:total_rot}), real-time estimated AoA values approach to truth AoA as the number of rotation goes up. In ideal setting where there is no reverberation (Setting 1), only two rotations were needed, two seconds, to get all correct AoA values in both $K=3,4$ cases. As channel becomes reverberant or noisy, more number of steps provide benefit in estimating true AoA values. Also, having longer $T_{unit}$ seem to help with identifying major peaks $\delta_k$ at initial step, but it does not help with getting rid of intrinsic convergence offset caused by reverberations.

\begin{figure}[hbt]
 
\centering \includegraphics[width=\columnwidth]{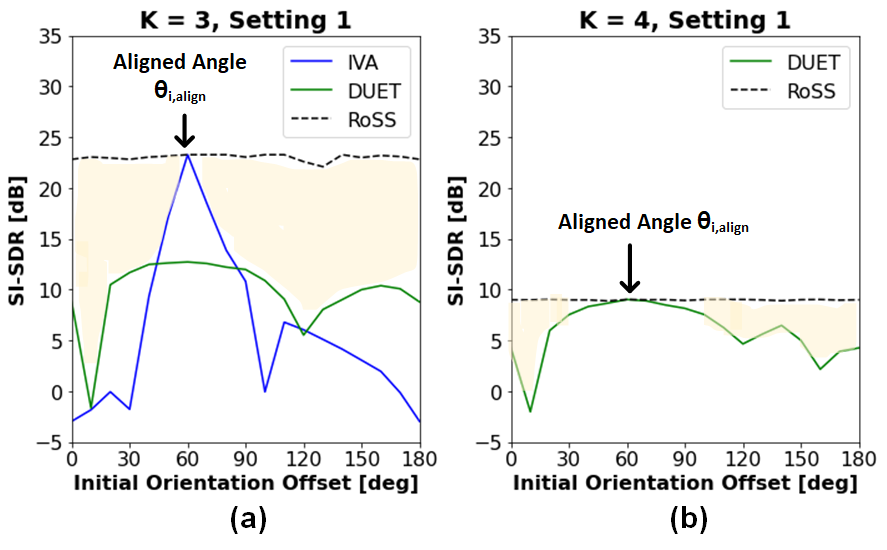}
  
\caption{\small{Source separation performance with initial orientation offset. Zero offset at (a) [170, 70, -50], (b) [170, 70, -50, -120].}}\label{fig:Start_Angle}
  
\end{figure}

\new
\noindent {\bf Performance Metric:}
For source separation metric, we utilize scale-invariant signal-to-distortion ratio (SI-SDR) and SI-SDRi \cite{Def_SISDR,TrainigNoisy} where SI-SDRi captures improvement on SI-SDR after source separation: $[$SI-SDR(estimate, target) - SI-SDR(mixture, target)$]$. In this work, SI-SDRi is used to deal with noisy outdoor data where both mixture data and target data are noisy due to wind.

\new
\noindent{\bf Benefit of RoSS:} Let us validate the effectiveness of interference alignment by considering the impact of initial orientation in terms of source separation. While AoAs of K sources are typically randomly distributed, conventional schemes such as IVA and DUET have huge variations in its performance depending on microphone array orientation in under-determined scenario (Fig. \ref{fig:Start_Angle}). And as expected, source separation is optimum at proposed alignment angles (\ref{eq:alignment_ang_K3})(\ref{eq:alignment_ang_K4}), each resulting in local AoA of [110, 10, -110] and [110, 10, -110, 180] when target was at 70 deg with zero offset. That is, in {\algo}, mobility provides an opportunity to rotate to that optimal angle point \emph{regardless of the initial orientation or source AoA distribution}. Therefore, filled area in yellow between RoSS (dotted line) and other methods represents the statistical gain of {\algo} in source separation. 

\begin{figure}[hbt]
 
\centering \includegraphics[width=\columnwidth]{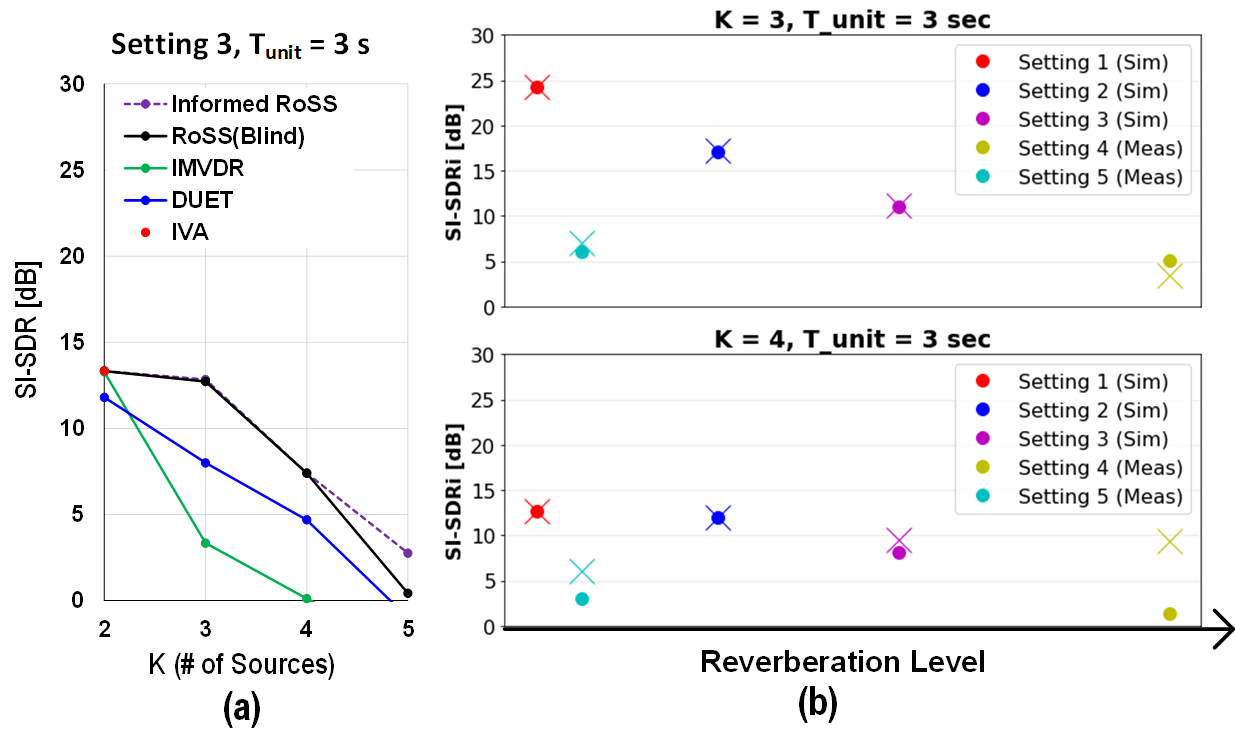}
  
\caption{\small{Average SI-SDRi compared to (a) various algorithms (b) different setups where X markers are AoA-informed RoSS.}}\label{fig:Reverb}
  
\end{figure} 

\new
\noindent{\bf Source Separation:} We show evaluation results on average SI-SDRi on each source for different settings for $K=3, 4, 5$ when $T_{unit}=3$ seconds in Fig. \ref{fig:Reverb}. In (a), about 1 degree of freedom gain is clear from AoA-informed RoSS and RoSS over other algorithms. In (b), for setting 1-to-4, as environment becomes more reverberant, SI-SDRi values drop because both AoA estimation and interference alignment starts to fail. However, setting 5 result shows poor performance even if it has almost no reverberation since it's outdoor measurement at the field. Major cause of this is thought to be external noises such as constant wind effect. 


 \section{Conclusion and Future Works}
In this paper, we have presented RoSS where it blindly searches AoAs via rotational motion to align a pair of interference signals to separate target sound source in under-determined scenario. While we demonstrated the concept with unsupervised blind schemes assuming sources are mutually independent, proposed idea of reducing the gap between $K$ and $M$ is applicable to many other UBSS/BSS method regardless of whether it's supervised or unsupervised even when sources are partially correlated. Also, we believe that applications where AoAs or source locations are readily-available, i.e. audio-vision applications, can benefit directly from proposed idea to align the interference as demonstrated in Fig. \ref{fig:Reverb} with informed RoSS bypassing the AoA estimation step to gain robustness against high K and reverberation.  

\new

As a next step, further investigation can be made on exploiting continuous motion since we currently consider only discrete rotations. Also, real-time score of recovered target source can be utilized to find the alignment angle.

\minus

\bibliographystyle{IEEEtran}
\bibliography{RoSS_IROS}

\end{document}